\documentclass[reprint, amsmath,amssymb,aps]{revtex4-1}
\usepackage{float}
\usepackage{siunitx}
\usepackage{placeins}
\usepackage{comment}
\usepackage{caption}
\usepackage{latexsym}
\usepackage{amssymb}
\usepackage{amsmath}
\usepackage{subcaption} 
\usepackage{graphicx} 
\usepackage{xcolor}
\usepackage{dcolumn}
\usepackage{bm}
\begin{document}
\title{Coalescence of elastic blisters filled with a viscous fluid}
\author{Torstein Sæter$^1$}
\author{Christian Pedersen$^1$}
\author{Jacco H. Snoeijer$^2$}
\author{Thomas Salez$^3$}
\email{thomas.salez@cnrs.fr}
\author{Andreas Carlson$^1$}
\email{acarlson@math.uio.no}
\affiliation{$^1$
Department of Mathematics, Mechanics Division, University of Oslo, Oslo 0316, Norway.}
\affiliation{$^2$
Physics of Fluids Group, Faculty of Science and Technology, University of Twente, 7500 AE Enschede, The Netherlands.}
\affiliation{$^3$
Univ. Bordeaux, CNRS, LOMA, UMR 5798, F-33400 Talence, France.}
\date{\today}
\begin{abstract}
Pockets of viscous fluid coalescing beneath an elastic plate are encountered in a wide range of natural phenomena and engineering processes, spanning across scales. As the pockets merge, a bridge is formed with a height increasing as the plate relaxes. We study the spatiotemporal dynamics of such an elasto-hydrodynamic coalescence process by combining experiments, lubrication theory and numerical simulations. The bridge height exhibits an exponential growth with time, which corresponds to a self-similar solution of the bending-driven thin-film equation.  We address this unique self-similarity and the self-similar shape of the bridge, both of which are corroborated in numerical simulations and experiments. 
\end{abstract}
\maketitle

Viscous flows beneath an elastic sheet hold significant relevance in various natural phenomena and industrial processes. Examples include flow-driven intrusions of elastic fronts \cite{hosoi2004peeling, Lister2013, Hewitt2015, Ball2018, PhysRevFluids.4.124003,lister_skinner_large_2019,Peng2020, michaut2019}, viscous adhesion of elastic sheets and cell membranes \cite{cells, poulain_carlson_mandre_mahadevan_2022}, soft viscous fingering instabilities \cite{pihler,part22015, JUEL2018}, as well as in geological processes such as sill or laccolith formation \cite{Kerr,Michaut2011, BungerCruden2011}. 

When two fluid pockets are trapped between an elastic sheet and a pre-wetted solid substrate, forming what we refer to as blisters, they will spread and eventually merge if they are in close proximity. As they meet, the elastic bending of the sheet will, at short-times, relax the system into a single blister that, at long times, flattens by spreading \cite{Lister2013,carlson2018fluctuation}. This situation is analogous to the extensively studied phenomenon of capillary-driven drop coalescence, where different flow regimes \cite{eggers_lister_stone_1999, duchemin_eggers_josserand_2003, Ristenpart2006, Hernandez2012} and the effect of liquid rheology \cite{ViscoplasticCoalescence, ViscoElasticCoalescence} have been characterized. Hern\'andez-S\'anchez \textit{et al.} \cite{Hernandez2012} demonstrated that the bridge connecting two coalescing viscous sessile droplets grows linearly with time, which was described by a similarity solution of the governing lubrication model. Recently, also the self-similar form for the three-dimensional bridge shape has been derived \cite{3dStone}. Despite the prevalence of elastohydrodynamic coalescence, a comprehensive understanding of the underlying physical processes involved is currently lacking. This Letter addresses this knowledge gap by employing a combination of experiments, lubrication theory, and numerical simulations. We reveal that the short-time asymptotic relaxation dynamics can be described by a universal self-similar solution with an anomalous exponential growth with time, and we quantify the details of the spatiotemporal dynamics. 
\begin{figure*}[!htbp]
\includegraphics[scale = 0.35]{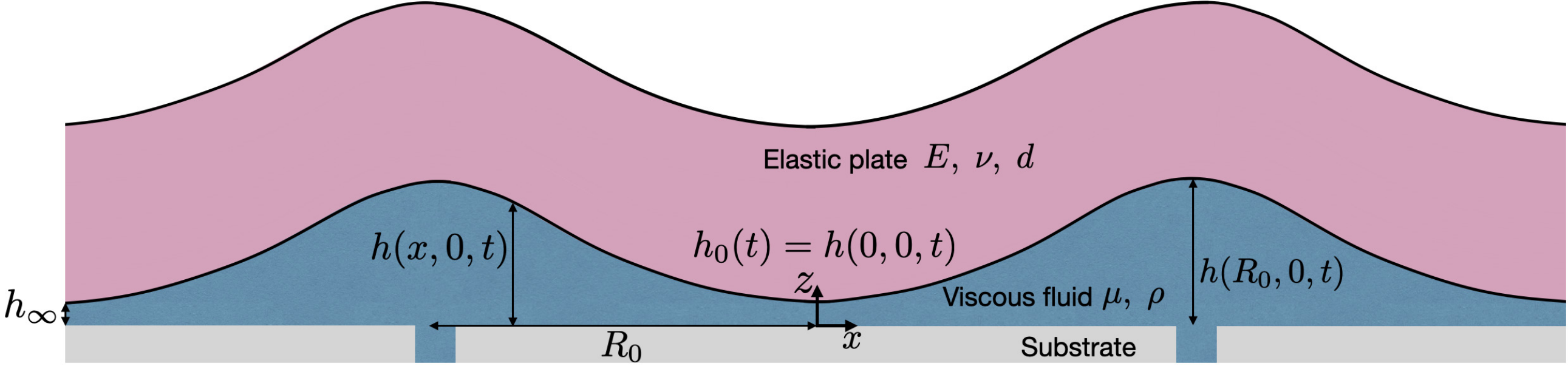}
\caption{\label{Fig1} A cross sectional schematic ($y=0$) of the studied system. Two identical pockets filled with a viscous fluid of viscosity $\mu$ and density $\rho$ are created by injecting prescribed fluid volumes beneath an elastic plate. The elastic plate has a Young modulus $E$, a Poisson ratio $\nu$, and a thickness $d$. After the fluid injection is stopped, each blister spreads until reaching a radius $R_0$ when they meet, which is defined as time $t=0$. The fluid height profile is $h(x,y,t)$, while $h_0(t) = h(0,0,t)$ is the bridge height, $h_{\textrm{i}} = h(\pm R_0,0,0)$ is the initial blister height, and $r(t)$ is the half-width of the bridge along the $y$-axis.}
\end{figure*}

We consider an elastic plate that is separated from the solid substrate by a thin liquid film of height $h_{\infty}$, as shown in Fig.~\ref{Fig1}. Two identical pockets, forming the blisters, are generated by the influx of viscous liquid through two inlets (diameters $4\cdot 10^{-3}$ \si{\meter}) at the supporting substrate, separated by a distance $2 R_0 = 0.15$ \si{\meter}. We define $t=0$ as the time the blisters make contact, with the height of the blister peak $h_{\textrm{i}}$. The system is designed to have $h_{\textrm{i}}/d<1$, where $d$ is the plate thickness, and $h_{\textrm{i}}/R_0\ll1$. The cross-sectional height profile $h(x,y=0,t)$ of the fluid film is obtained by passing a laser line along the $x$-coordinate through the blister peaks. A Nikon camera captures images of the laser line every $2$ \si{\second}, at an angle of $25^\circ$ relative to the horizontal plane. The measured laser line is fitted with a Gaussian intensity distribution along each vertical column of pixels. This imaging setup allows for the visualization and measurement of the height profile with time. 

The circular elastic plate of diameter $D=0.4$ \si{\meter} and thickness $d = 5.7\cdot 10^{-3} $ \si{\meter} is made from a silicon-based elastomer (Zhermack, Elite Double), with Young's modulus $E = 0.25$ MPa and Poisson ratio $\nu = 0.5$ \cite{ED}, giving a bending stiffness $B=Ed^3/[12(1-\nu^2)]=5\cdot 10^{-3}$ \si{\newton \meter}. Balancing elastic bending and gravity, we find the elastogravity length $[B/(\rho g)]^{1/4} = 28 \cdot 10^{-3}$ \si{\meter} $= 0.36\, R_0$, with $g$ the gravitational acceleration and $\rho$ the fluid density. Therefore, we must take into account the influence of gravity \cite{Lister2013, michaut2019, PhysRevFluids.4.124003}. Different silicone oils, with viscosities $\mu \in [0.1, 0.35, 0.5]$ \si{\pascal \second} and density $\rho=970$ kg m$^{-3}$, are used as fluids. The supporting rigid plexiglass substrate and the bottom side of the flat elastic plate are both precoated with oil using a squeegee, before gently put in contact.  A fluid film of thickness $h_{\infty} = (40 \pm 14)\cdot 10^{-6}$ \si{\meter} hence separates the two flat solids before any fluid injection through the inlets. $h_{\infty} $ is determined on a smaller plate ($21 \times 24\cdot 10^{-4}$ \si{\metre \squared}) by weighing it with and without the prewetted film, with an error estimate determined from 40 repetitions. Air bubbles trapped underneath the elastic plate are visually identified through the transparent substrate. These bubbles are removed by squeezing the elastic plate in the corresponding regions, transporting the bubbles to the edge. 

After injecting a prescribed fluid volume with a Merck-Millipore-Sigma syringe pump, the two blisters spread until they make contact. The experimental determination of the contact point is partly limited by our spatial resolution $0.067$ \textrm{mm/pixel}, leading to an uncertainty in defining the contact time $t=0$. It is also worth noting that a height difference of up to $10\%$ between the two blister peaks can occur. We thus define $h_{\textrm{i}}$ as the average between the two blister heights at $t=0$. Varying the injected fluid volume, we obtain $h_{\textrm{i}}\in[1.9-4.2]\cdot 10^{-3}$ \si{\meter}.  

We use the elastohydrodynamic lubrication theory \cite{oron1997long, hosoi2004peeling, craster2009dynamics, pedersen2021universal} to rationalize the experimental measurements. Assuming no-slip at the two solid boundaries, conservation of mass, and an incompressible Newtonian lubrication flow, the fluid height profile $h(x,y,t)$ satisfies the thin-film equation:
\begin{equation}
\frac{\partial h(x,y,t)}{\partial t } = \frac{1}{12\mu}\nabla \cdot \left[ h^3(x,y,t)\nabla p(x,y,t)\right],
\label{eq:lub}
\end{equation}
\begin{equation}
p(x,y,t) = B \nabla^4 h(x,y,t) + \rho g [h(x,y,t)-h_{\infty}],
\label{eq:pressure}
\end{equation}
where $\nabla = \left(\frac{\partial }{\partial x},\frac{\partial }{\partial y}\right)$ is the nabla operator. $p(x,y,t)$ is the excess hydrodynamic pressure generated by elastic bending of the plate and gravity, neglecting any effect from stretching the plate as the edge is free to move and $h_{\textrm{i}}<d$ \cite{michaut2019, PhysRevFluids.4.124003}. To cast Eqs. ~(\eqref{eq:lub}, \eqref{eq:pressure}) and our data in dimensionless units $\tilde{(~)}$, we use the scaling relations: $\tilde{x}=x/R_0$, $\tilde{y}=y/R_0$, $\tilde{h}(\tilde{x},\tilde{y},\tilde{t})=h(x,y,t)/h_{\textrm{i}}$, $\tilde{h}_{\infty} = h_{\infty}/h_{\textrm{i}}$, and $\tilde t=t/T$ with $T = \mu R_0^6/(B h_{\textrm{i}}^3)$. Inserting these in Eqs.~(\eqref{eq:lub}, \eqref{eq:pressure}) yields the dimensionless thin-film equation, with a single dimensionless number $N = \rho g R_0^4/B$ in front of the gravity term, which is the ratio between gravity and elastic bending, \textit{i.e.} an elastic equivalent of the Bond number. 

Eqs.~(\ref{eq:lub},~\ref{eq:pressure}) are solved numerically by using a finite-element method. We discretize the equations using linear elements and use an implicit-time marching scheme \citep{PhysRevFluids.4.124003}. At $\tilde{t}^*=0$, the simulations are initialized with two identical bumps centered at $(\tilde{x},\tilde{y}) = (\pm 1, 0)$, as: $\tilde{h}(\tilde{x},\tilde{y},\tilde{t}^*=0) = \tilde{h}_{\infty} + 0.6\left[\left(1-\frac{(\tilde{x}\pm 1)^2 + \tilde{y}^2}{0.75^2}\right)^2\right]$ for $\sqrt{(\tilde{x} \pm 1)^2 +\tilde{y}^2} \leq 0.75$, and $\tilde{h}(\tilde{x},\tilde{y},\tilde{t}^*=0) = \tilde{h}_{\infty}$ otherwise. After a few time steps, the height profile becomes smooth and the two blisters spread a short distance before they make contact. We define the moment of contact as the new temporal origin $\tilde{t}=0$, and rescale the profile so that $\tilde{h}(\pm 1,0,0) = 1$. To reduce the computational time, we only simulate a quarter of each of blister, and therefore apply symmetry conditions: $\nabla \tilde{h}(x,y,t)\cdot\mathbf{n} = \nabla^3 \tilde{h}(x,y,t)\cdot\mathbf{n} = \nabla^5 \tilde{h}(x,y,t)\cdot\mathbf{n} = 0$ at all boundaries, with $\mathbf{n}$ their normal vector. We use a non-uniform grid, which is refined around the bridge with minimum mesh size $\Delta \tilde{x} = 1.25\cdot 10^{-4}$, and an adaptive time stepping routine with maximum time step $\Delta \tilde{t} = 1.6\cdot 10^{-4}$.
\begin{figure}[!htbp]
\centering
\includegraphics[scale = 0.34]{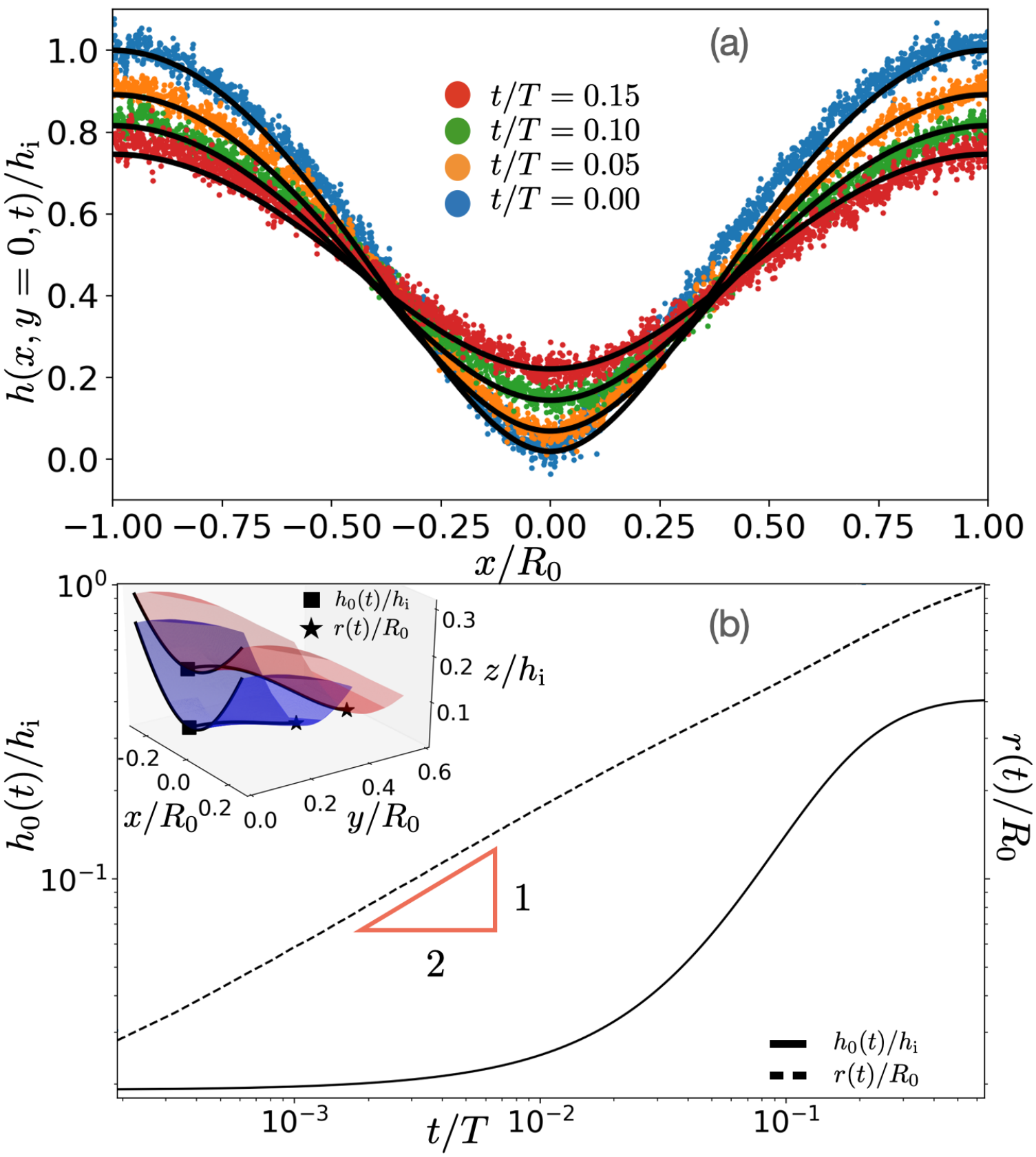}
\caption{\label{fig:heightprofiles} 
(a) Dimensionless cross-sectional height profiles $h(x,y=0,t)/h_{\textrm{i}}$, at different dimensionless times $t/T$, from an experiment with $\mu = 0.5$ \textrm{Pa s} and $h_{\textrm{i}} = 2.55$ \si{\milli \meter}, and from the numerical solutions (solid lines) of Eq.~\eqref{eq:lub} with $N=59$ and $h_{\infty}/h_{\textrm{i}} =0.019$. (b) Dimensionless height $h_0(t)/h_{\textrm{i}}$, and half-width $r(t)/R_0$ of the bridge, extracted from the numerical solutions of Eq.~\eqref{eq:lub}. Inset: Dimensionless height profiles $h(x,y,t)/h_{\textrm{i}}$ from the numerical simulation of Eq.~\eqref{eq:lub}, at dimensionless times $t/T = 0.11$ (blue) and $t/T =0.19$ (red). The bridge half-width $r(t)$ and height $h_0(t)$ are indicated by the star and square markers, respectively.}
\end{figure} 

We now turn our attention to the coalescence dynamics, after the blisters have made contact. An example of temporal evolution of the height profiles is illustrated in Fig.~\ref{fig:heightprofiles}(a). The numerical solutions and the experiments are in close agreement. The thickness $h_{\infty}=40\cdot 10^{-6}$ \si{\meter} of the prewetted layer together with the peak height $h_{\textrm{i}} = 2.55 \cdot 10^{-3}$ \si{\meter}, obtained from the experiments, give $\tilde{h}_{\infty} = 0.016$, similar to the value $\tilde{h}_{\infty} = 0.019$ used in the numerical simulation. From the numerical height profiles, we extract the bridge half-width $r(t)$ and height $h_0(t)=h(0,0,t)$, as shown in Fig.~\ref{fig:heightprofiles}(b). The half-width is defined as the first point along $y$ where $h(x=0,y>0,t)\leq h_{\infty}$. It is striking to observe that these two quantities ($r(t),h_0(t)$) have very different behaviour with time: the half-width follows a power law $r(t)\sim t^{1/2}$, which is reminiscent to the capillary case \cite{Ristenpart2006}, while the bridge height $h_0(t)$ clearly deviates from a power law behaviour.

To understand the bridge height dynamics, we take inspiration from the self-similar analysis of the elastohydrodynamic blister growth \cite{Lister2013}, and from the capillary coalescence case \cite{Ristenpart2006, Hernandez2012}. We assume that the flow is predominantly oriented along the $x$-direction, so that we can consider a two-dimensional problem. At the small scales associated with the initial stages of coalescence, the dynamics is driven by elastic bending and gravity can be neglected. Therefore, the dimensionless version of Eq.~\eqref{eq:lub} simplifies to:
\begin{equation}
\frac{\partial \tilde{h}(\tilde x,\tilde t)}{\partial \tilde{t} } = \frac{1}{12}\frac{\partial}{\partial \tilde{x}}\left[ \tilde{h}^3(\tilde x,\tilde t)\frac{\partial^5 \tilde{h}(\tilde x,\tilde t)}{\partial \tilde{x}^5} \right].
\label{eq:non-dim}
\end{equation} 
\begin{figure*}[ht]
\includegraphics[scale = 0.35]{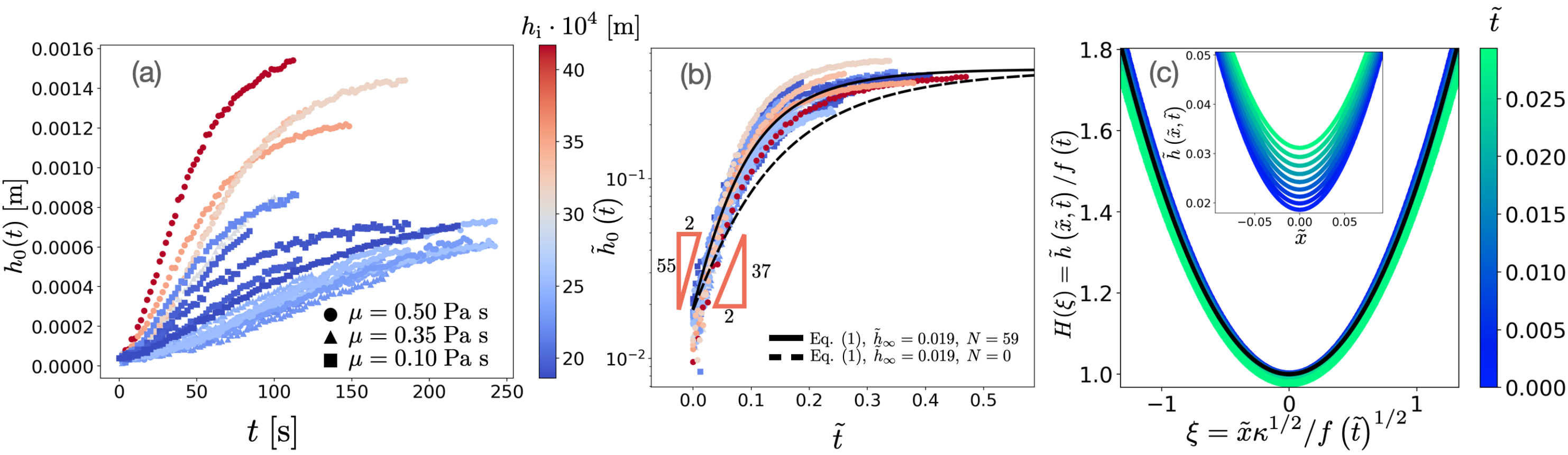}
\caption{\label{fig:exp} (a) Experimental bridge height $h_0(t)$ for different viscosities
$\mu \in [0.1, 0.35, 0.5]$ Pa s and  blister peak heights $h_{\textrm{i}}\in[1.9-4.2]\cdot 10^{-3}$\si{\meter}. (b) Experimental dimensionless bridge height $\tilde{h}_0(\tilde{t}) = h_0(\tilde{t})/h_{\textrm{i}}$ as a function of dimensionless time $\tilde{t}=t/T$. The solid and dashed lines represent the numerical solutions of Eq.~\eqref{eq:lub} for $N=59$ and $N=0$, respectively, and with $\tilde{h}_{\infty}=0.019$. The short-time behaviour for $N=59$ is well described by $\tilde{h}_0(\tilde{t}) =  \tilde{h}_0(0)\exp\left(27.5\tilde{t}\right)$ ; while for $N=0$ it follows $\tilde{h}_0(\tilde{t}) =  \tilde{h}_0(0)\exp\left(18.5\tilde{t}\right)$. (c) Inset: The height profiles $\tilde{h}(\tilde{x},\tilde{y}=0,\tilde{t})$ (noted $\tilde{h}(\tilde{x},\tilde{t})$) from the short-time numerical solution of Eq.~\eqref{eq:lub} for $N=0$ and $\tilde{h}_{\infty}=0.019$. Main: Self-similar representation of the height profiles ($\tilde{y}=0$) from Eq.~\eqref{eq:lub} shown in the inset. Here, we used $f(\tilde{t}) =  1.9\cdot 10^{-3}\exp \left (18.5\tilde{t}\right)$ and $\kappa = 8$. The black solid line is the numerical solution of Eq.~\eqref{eq:ODE} with $\kappa = 8$ and $\alpha = 0.0181$.}  
\end{figure*}

Let us look for a similarity solution $H(\xi)=\tilde{h}( \tilde{x},\tilde{t})/f(\tilde{t})$ of Eq.~(\ref{eq:non-dim}), with $\xi = \tilde{x}/g(\tilde{t})$, where $f(\tilde{t})$ and $g(\tilde{t})$ are two unknown functions~\cite{barenblatt_1996}. At short-times, we consider the quasi-static shape of the blister~\cite{Lister2013} as a constant outer solution $\xi\rightarrow\infty$ for the coalescence dynamics, instead of the interface slope (or contact angle) used for the capillary case \cite{Hernandez2012}. Since the quasi-static blister shape has a non-zero dimensionless ``contact curvature" $\kappa$~\cite{Lister2013}, we impose here that: 
\begin{equation}
\frac{f(\tilde{t})}{g^2(\tilde{t})}H''(\infty) = \kappa.
\label{eq:curvature}
\end{equation}
For the case without gravity ($N=0$), the contact curvature follows from the blister shape at constant bending pressure~\cite{Lister2013}, which in the present units gives $\kappa_0=8$. For the case with gravity ($N=59$), the contact curvature $\kappa=8.8$ is found to be slightly larger. For convenience, and without loss of generality, we normalise the similarity solution such that $H''(\infty)=1$, from which  it follows that $f(\tilde{t})=\kappa g^2(\tilde{t})$. Inserting the similarity variable transform into Eq.~\eqref{eq:non-dim}, we obtain:
\begin{equation}
\frac{1}{\kappa^3}\frac{\dot{g}\left(\tilde{t}\right)}{g\left(\tilde{t}\right)} = \alpha =\frac{1}{12}\,\frac{\left[H \left(\xi\right)^3H'''''\left(\xi\right)\right]'}{2H \left(\xi\right) - \xi H'\left(\xi\right)},
\label{eq:ODE}
\end{equation}
where the dot and prime indicate derivatives with respect to $\tilde{t}$ and $\xi$, respectively. Owing to the separation of variables, the parameter $\alpha$ must be a constant that needs to be determined. We thus obtain two ordinary differential equations (ODEs). The solution of the ODE for $g(\tilde{t})$ reads:
\begin{equation}
g(\tilde{t}) = g(0) \exp \left( \beta \tilde{t}\right), \quad \mathrm{with} \quad  \beta = \alpha \kappa^3.
\label{eq:exp}
\end{equation}
As a consequence, one gets $f(\tilde{t})=\kappa g(0)^2 \exp \left(2\beta \tilde{t}\right)$. To the best of our knowledge, this is the first time that an anomalous exponential self-similar solution is found in the context of lubrication flow. Such self-similar solutions have, however, been found in other physical phenomena, \textit{e.g.} in diffusion dynamics \cite{Iagar_2022}. 

We proceed to test the prediction from Eq.~(\ref{eq:exp}) by using the experimental and numerical data. The experimental data of the temporal evolution of the bridge height $h_0(t)$, for $h_{\textrm{i}} \in [1.9-4.2]\cdot 10^{-3}$ \si{\meter} and $\mu \in [0.1,0.35,0.5]$ Pa s, is shown in Fig.~\ref{fig:exp}(a). It is clear that both $\mu$ and $h_{\textrm{i}}$ affect the relaxation dynamics. As expected, increasing the viscosity slows down the dynamics. The effect of $h_{\textrm{i}}$ is more subtle, but consistent with the prediction $\beta=\alpha\kappa^3$, as a larger $h_{\textrm{i}}$ increases the contact curvature of the blisters, thereby enhancing the growth rate. In Fig.~\ref{fig:exp}(b), the same data is recast into dimensionless form that collapses it onto a nearly single curve. The numerical solution of Eq.~\eqref{eq:lub} with $N=59$ reproduces the dynamics measured in the experiments, which at short-times corresponds to an exponential growth with $\beta=13.75$. For comparison, the numerical solution of Eq.~\eqref{eq:lub} with $N=0$ is also shown, which at short-times has a $\beta_0=9.25$. Assuming that $\alpha$ is independent of $N$, one gets that $\beta/\beta_0$ is given by $(\kappa/\kappa_0)^3$, as observed within a $\sim12\%$ margin of error. The error is likely stemming from uncertainties in the exponential fits and in the extracted values for $\kappa$ and $\kappa_0$. However, $\alpha$ does depend on the dimensionless prewetted layer thickness $\tilde{h}_{\infty}$, where a smaller $\tilde{h}_{\infty}$ leads to a smaller exponential growth rate (not shown). Therefore, $\alpha$ is not a universal constant, and equals $\alpha=\beta_0/\kappa_0^3\approx0.0181$ in our case.

Finally, we investigate the self-similarity of the bridge profile $H (\xi)$ through Eq.~(\ref{eq:ODE}). The latter is a sixth-order nonlinear ODE, which is solved for $\alpha= 0.0181$, using the numerical solver \textit{bvp5c} in MATLAB with boundary conditions: $H'(0)= H'''(0) = H'''''(0) = 0$, $H''(\infty) = 1$, and $H'''(\infty) = H''''(\infty) = 0$. Since the bridge is symmetric around $\xi=0$ and has a constant far-field curvature it provides the required six boundary conditions, so we do not impose $H(0) = 1$ in contrast to the capillary case~\cite{Hernandez2012}. Nevertheless, by introducing the rescaled variables $\xi^* = \xi/H(0)^{1/2}$ and $H^*(\xi^*) = H(\xi)/H(0)$, one can always enforce $H^*(0)=1$, without having it explicitly stated in the boundary conditions. The results for the self-similar profile are shown in Fig.~\ref{fig:exp}(c). The close agreement between the numerical solution of Eq.~(\ref{eq:ODE}) and the numerical solution of Eq.~(\ref{eq:lub}) for $N=0$, corroborate the validity of the similarity solution. 

In this Letter, we have demonstrated that, as two pockets of viscous fluids merge under an elastic plate, the connecting bridge has a height that grows exponentially with time. The spatio-temporal dynamics agrees between the experiments and the numerical solutions of the thin-film equation. Moreover, a self-similar exponential solution is found, which rationalises the short-time dynamics found in the experiments and numerical simulations. To the best of our knowledge, this represents the first exponential self-similar solution for lubrication flows, standing in sharp contrast to the conventional power-law solutions. Our finding highlights the significant and non-trivial role played by the interaction between elastic deformations and viscous lubrication flows. 
\\

The authors gratefully thank Federico Hern\'andez-S\'anchez for interesting discussions. They acknowledge financial support from the EarthFlow initiative at the University of Oslo. They also acknowledge financial support from the European Union through the European Research Council under EMetBrown (ERC-CoG-101039103) grant. Views and opinions expressed are however those of the authors only and do not necessarily reflect those of the European Union or the European Research Council. Neither the European Union nor the granting authority can be held responsible for them. Besides, the authors acknowledge financial support from the Agence Nationale de la Recherche under the EMetBrown (ANR-21-ERCC-0010-01), Softer (ANR-21-CE06-0029), and Fricolas (ANR-21-CE06-0039) grants. Finally, they thank the Soft Matter Collaborative Research Unit, Frontier Research Center for Advanced Material and Life Science, Faculty of Advanced Life Science at Hokkaido University, Sapporo, Japan.  

\bibliography{Saeter2023.bib}
\end{document}